\documentclass[12pt,preprint]{aastex}
\usepackage{rotating}
\usepackage{epsfig}
\usepackage{ifpdf}

\begin{document}

\title{JVLA Observations of IC 348SW: Compact Radio Sources and their Nature}

\author{Luis F. Rodr\'\i guez\altaffilmark{1,2}, Luis A. Zapata\altaffilmark{1} and Aina Palau\altaffilmark{1}}

\altaffiltext{1}{Centro de Radioastronom\'\i a y Astrof\'\i sica, 
UNAM, Apdo. Postal 3-72 (Xangari), 58089 Morelia, Michoac\'an, M\'exico}

\altaffiltext{2}{Astronomy Department, Faculty of Science, King Abdulaziz University, 
P.O. Box 80203, Jeddah 21589, Saudi Arabia}

\email{l.rodriguez,l.zapata,a.palau@crya.unam.mx}
 
\begin{abstract}
We present sensitive 2.1 and 3.3 cm JVLA radio continuum observations of the region IC 348 SW.
We detect a total of 10 compact radio sources in the region, of
which seven are first reported here. One of the sources is associated
with the remarkable periodic time-variable infrared source LRLL 54361, opening the
possibility of monitoring this object at radio wavelengths.
Four of the sources appear to be powering outflows in the region, including HH 211 and
HH 797. In the case of the rotating outflow HH 797 we detect at its center a double radio source, separated
by $\sim3''$. Two of the sources 
are associated with infrared stars that possibly have gyrosynchrotron emission
produced in active magnetospheres. Finally, three of the sources are interpreted as background objects. 

\end{abstract}  


\section{Introduction}

IC 348 is a young ($\sim$2-3~Myr) open cluster with more than 300
members identified from optical and infrared observations (Herbig 1998; Lada \& Lada 1995;
Preibisch 2003). Its distance is usually estimated to be that of the Perseus molecular cloud complex,
$\sim$300 pc (Herbig 1998), although more recent maser parallax determinations
to NGC1333 and L1448 favor $\sim$240 pc (Hirota et al. 2008; 2011, see also Chen et al.
2013). The cluster extends about $15' \times 15'$ in the sky,
approximately centered on the B5 V star BD +31$^\circ$ 643,
the most massive and optically brightest member of the cluster.
The southwest part of the cluster (that we refer to as IC 348SW) is particularly
interesting because it hosts two highly collimated Herbig-Haro (HH) outflows: HH 211 and HH 797.

HH 797 is one of the best cases of an apparently rotating molecular outflow (Pech et al. 2012).
These are systems that present a velocity gradient perpendicular to the flow axis.
This gradient has been interpreted as produced by rotation in the flow, although several
other explanations have been advanced (Soker 2005). 
One alternative explanation for the apparent rotation is that we are dealing
with a double outflow system, with the outflows nearly parallel but unresolved spatially.
The exciting sources of outflows are usually detected as weak free-free centimeter sources,
with this emission produced by the ionized jets close to the star (e.g. Rodr\'\i guez \&
Reipurth 1998). Then, the detection of a double radio source at the center of
the outflow system would support the interpretation of a double outflow system. In this paper
we present sensitive Jansky Very Large Array (JVLA) observations of IC 348SW that may address this issue.

\section{Observations}
The observations were made with the Karl G. Jansky Very Large Array of NRAO\footnote{The National 
Radio Astronomy Observatory is a facility of the National Science Foundation operated
under cooperative agreement by Associated Universities, Inc.} centered at rest frequencies of 14.0 (2.1 cm) and 9.0 
(3.3 cm) GHz during
2013 June.  At that time the array was in its C configuration.  The phase center was 
at $\alpha(2000) = 03^h~ 43^m~ 57\rlap.^s0$;
$\delta(2000)$ = $+$32$^\circ~ 03'~ 04.0''$. For both observations the absolute amplitude calibrator was J1337$+$3309 and
the phase calibrator was J0336$+$3218. 

The digital correlator of the JVLA was configured in 31 spectral windows of 128 MHz width divided 
in 64 channels of spectral resolution. The first 15 windows were used to observe at 2.1 cm, and the rest to observe at 3.3 cm.
The total bandwidth for both observations was about 2.048 GHz in a dual-polarization mode.
The half power width of the primary beam is $5\rlap{'}.0$ at 3.3 cm and $3\rlap{'}.2$ at 2.1 cm. 

The data were analyzed in the standard manner using CASA (Common Astronomy Software Applications) package of NRAO,
although for some stages of the analysis we used the AIPS (Astronomical Image Processing System)
package. 
In both observations, we used the ROBUST parameter of CLEAN set to 2, to obtain a better sensitivity losing some 
angular resolution. 
To construct the continuum in both bands, we only used the free-line channels. 
At 2.1 cm, the resulting image rms was 5 $\mu$Jy beam$^{-1}$ at an angular 
resolution of $2\rlap.{''}40 \times 1\rlap.{''}67$ with PA = $-75\rlap.^\circ2$.
On the other hand, at 3.3 cm,  the resulting image rms was  4 $\mu$Jy beam$^{-1}$ at an angular resolution 
of $3\rlap.{''}92 \times 2\rlap.{''}52$ with PA = $-77\rlap.^\circ5$.

\begin{figure}
\centering
\includegraphics[angle=0,scale=0.45]{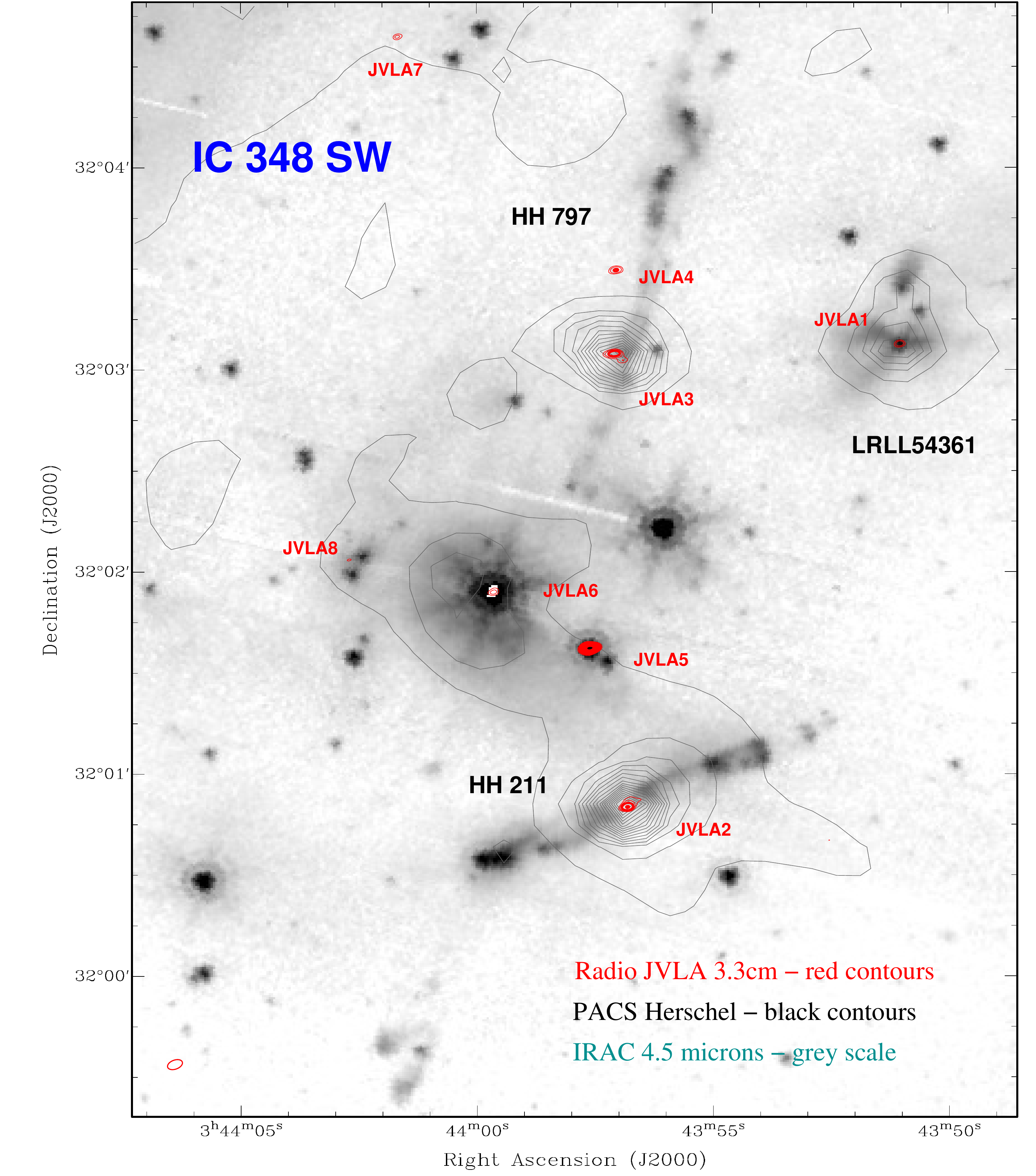}
\caption{\small JVLA 3.3 cm  (red) and PACS/Herschel 160 $\mu$m (black)  continuum contour images overlaid 
on a Spitzer IRAC 4.5 $\mu$m  grey-scale image from IC 348. 
The red contours are 24, 30, 40, 60, 80, 100,
and 120 $\mu$Jy beam$^{-1}$ and the rms noise of the
image is 4 $\mu$Jy beam$^{-1}$. 
The black contours range from 1\% to 86\% of the peak emission, in steps of 5\%. 
The emission peak for the far-infrared emission is 5.38 Jy beam$^{-1}$.
The half-power contour of the synthesized beam of the 3.3 cm image is shown in the bottom left corner.}
\label{fig1}
\end{figure}

\pagebreak

\begin{figure}
\centering
\vspace{-1.2cm}
\includegraphics[angle=0,scale=0.45]{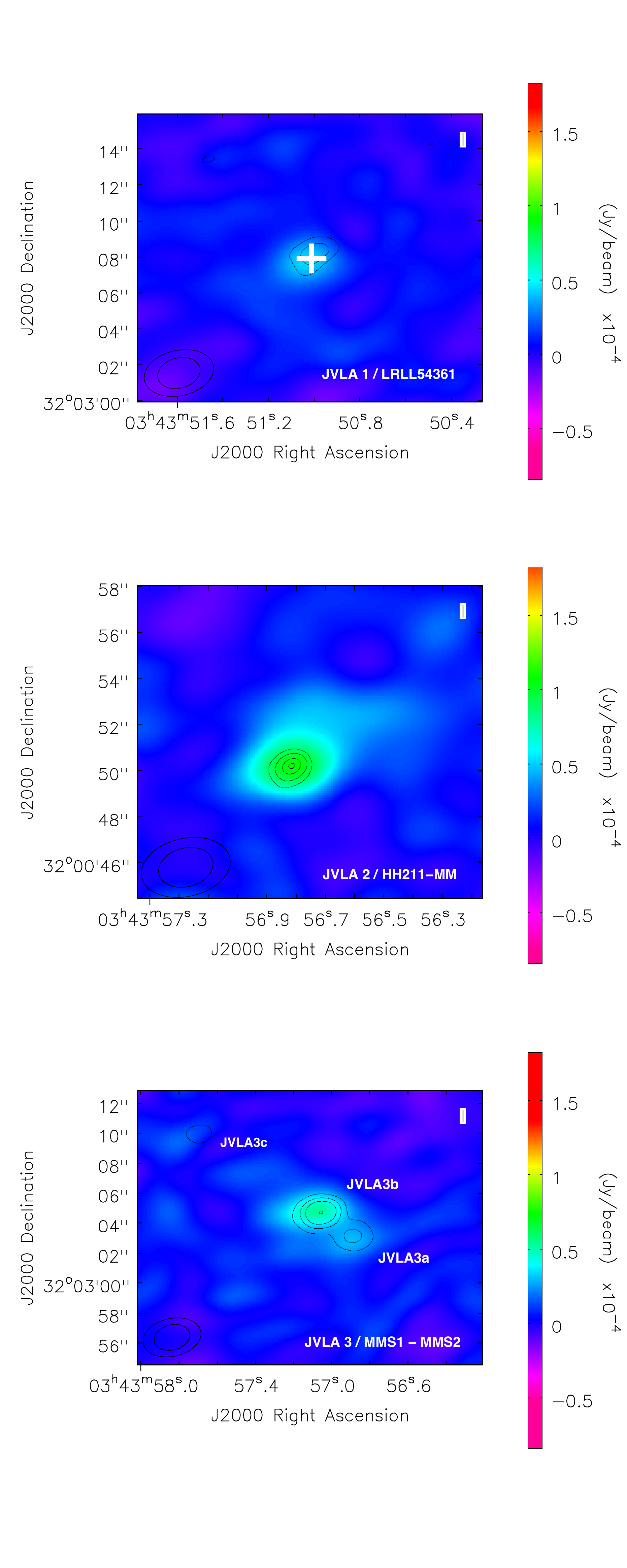}
\vskip-1.5cm
\caption{\small  JVLA 2.1 cm continuum contour image overlaid in a  JVLA 3.3 cm (color scale indicated
by the bar on the right side of the panels) image for
several sources in  IC 348 SW. In the upper panel,
the black contours are 20 and 37 $\mu$Jy beam$^{-1}$, in the middle panel are from  
81, 98, 116, and 125 $\mu$Jy beam$^{-1}$,
and in the lower panel are 20, 37, 54 and 87 $\mu$Jy beam$^{-1}$. The cross in the
upper panel marks the centroid position of the Spitzer/IRAC source (Muzerolle et al. 2013).
The half-power contour of the synthesized beams of the 3.3 and 2.1 cm images are shown in the bottom left corner.
The letter "I" in the upper right corner indicates the images are of the Stokes I parameter.}
\label{fig2}
\end{figure}

\pagebreak

\begin{deluxetable}{l c c c c c c c c c}
\tabletypesize{\scriptsize}
\tablecaption{Parameters of the JVLA sources detected at 3.3 cm}
\tablehead{                        
\colhead{}                        &
\colhead{}                        &
\multicolumn{2}{c}{Position} &
\colhead{}                              &
\multicolumn{3}{c}{Deconvolved size$^b$} &        \\
\colhead{}   &
\colhead{}   &
\colhead{$\alpha_{2000}$}          &
\colhead{$\delta_{2000}$}           &
\colhead{Flux Density$^a$ }       &                            
\colhead{Maj.}  &
\colhead{Min.}  &
\colhead{P.A.}  & \\
\colhead{Source}                              &
\colhead{Name}                              &
\colhead{(h m s) }                     &
\colhead{($^\circ$ $^{\prime}$  $^{\prime\prime}$)}              &
\colhead{($\mu$Jy)}  & 
\colhead{($^{\prime\prime}$)}  &
\colhead{($^{\prime\prime}$)}  &
\colhead{($^\circ$)} &
}
\startdata

JVLA 1     &   LRLL 54361  & 03 43 51.043 & $+$32 03 07.62  &  53 $\pm$ 11 &  - &  - & -\\
JVLA 2     &   HH 211-MM   & 03 43 56.781 & $+$32 00 50.77   &  186 $\pm$ 25 &  - &  - & - \\
JVLA 3a   &  HH 797-SMM2  & 03 43 56.928 & $+$32 03 03.10   &  47  $\pm$ 6 &  4.2 $\pm$ 1.2 &  2.2 $\pm$ 1.1 & 56 $\pm$ 38\\
JVLA 3b   &  HH 797-SMM2  & 03 43 57.093 & $+$32 03 04.69  & 79 $\pm$ 17 &  3.4 $\pm$ 0.9 &  0.6 $\pm$ 0.4 & 77 $\pm$ 14\\
JVLA 3c   &   HH 797-SMM2E      & 03 43 57.703 & $+$32 03 10.00   & $\leq$ 16 &  -&  - & -\\
JVLA 4     &    -   & 03 43 57.050 & $+$32 03 29.62   &   48 $\pm$ 11 &  2.0 $\pm$ 1.2 &  0.6 $\pm$ 0.5 & 93 $\pm$ 40\\
JVLA 5     &   IC 348 LRL 49          & 03 43 57.610 & $+$32 01 37.36   &  188 $\pm$ 15 &  -&  - & - \\
JVLA 6     &   IC 348 LRL 13          & 03 43 59.632 & $+$32 01 54.14 &  46 $\pm$ 5 &  -&  - & -\\
JVLA 7     &    -                               & 03 44 01.662 & $+$32 04 38.87   &   47 $\pm$ 7 &  -&  - & -\\
JVLA 8     &    -                               & 03 44 02.653 & $+$32 02 03.79 &  32 $\pm$ 5 &  -&  - & -\\
\enddata
\tablecomments{
                (a): Total flux density corrected for primary beam response,
obtained from the task JMFIT of AIPS. Upper limits are at the 4-$\sigma$ level.\\
                (b): These values were obtained from the task JMFIT of AIPS.}
\end{deluxetable}

\pagebreak

\begin{deluxetable}{l c c c c c c}
\tabletypesize{\scriptsize}
\tablecaption{Parameters of the JVLA sources detected at 2.1 cm}
\tablehead{                        
\colhead{}                        &
\multicolumn{2}{c}{Position} &
\colhead{}                              & \\
\colhead{}   &
\colhead{$\alpha_{2000}$}          &
\colhead{$\delta_{2000}$}           &
\colhead{Flux Density$^a$ }       &                            
\colhead{Spectral}   & \\
\colhead{Source}                              &
\colhead{(h m s) }                     &
\colhead{($^\circ$ $^{\prime}$  $^{\prime\prime}$)}              &
\colhead{($\mu$Jy)}  & 
\colhead{Index} &
}
\startdata

JVLA 1          & 03 43 51.001 & $+$32 03 08.16 &  49 $\pm$ 9 &  $-$0.2 $\pm$ 0.6$^b$\\
JVLA 2          & 03 43 56.819 & $+$32 00 50.10 &  210 $\pm$ 22 & $+$0.3 $\pm$ 0.4\\
JVLA 3a        & 03 43 56.892 & $+$32 03 03.21 &  65 $\pm$ 15 & $+$0.7 $\pm$ 0.6\\
JVLA 3b        & 03 43 57.053 & $+$32 03 04.64 &  104 $\pm$ 16 & $+$0.6 $\pm$ 0.6  \\
JVLA 3c        & 03 43 57.703 & $+$32 03 10.00 &  27 $\pm$ 8 & $\geq$1.2 $\pm$ 0.9 \\
JVLA 4          & 03 43 57.073 & $+$32 03 29.51 &   28 $\pm$ 9 & $-$1.2 $\pm$ 0.9 \\
JVLA 5          & 03 43 57.619 & $+$32 01 37.43 &  142 $\pm$ 14 & $-$0.6 $\pm$ 0.3 \\
JVLA 6          & 03 43 59.647 & $+$32 01 54.10 &  54 $\pm$ 8 &  $+$0.4 $\pm$ 0.4\\
JVLA 7          & 03 44 01.662 & $+$32 04 38.87   &  $\leq$ 20 & $\leq -$1.9 $\pm$ 0.7 \\
JVLA 8          & 03 44 02.653 & $+$32 02 03.79 & $\leq$ 20  & $\leq -$1.1 $\pm$ 0.7 \\

\enddata
\tablecomments{
                (a): Total flux density corrected for primary beam response,
obtained from the task JMFIT of AIPS. Upper limits are at the 4-$\sigma$ level.\\
                (b): The spectral index was obtained assuming that the sources detected at 3.3 and 2.1 cm are the same.}
\end{deluxetable}

\section{Comments on individual sources}

The radio sources with peak flux densities above 5-$\sigma$ at 3.3 and 2.1 cm 
were considered detections and are listed in Tables 1 and 2,
respectively. In Table 1 we give the JVLA number of the source, its possible association reported
in the literature, its right ascension and declination, its flux density and the deconvolved dimensions
of three of the sources. In Table 2 we give the JVLA number of the source, its right ascension and declination, 
its flux density and the derived spectral index. All sources at 2.1 cm appeared as unresolved and no
deconvolved dimensions are presented in Table 2. 
Of the total of 10 objects reported here, only three (sources JVLA 2, 3b, and 5) had been reported
previously as radio sources.

\subsection{JVLA1}

This radio source is associated with the infrared source LRLL 54361 (Luhman et al. 1998), as can be seen in 
Figure 1. In this figure we show the 3.3 cm JVLA emission as well as the PACS/Herschel 160 $\mu$m
and IRAC 4.5 $\mu$m emissions. 
Recently, the infrared luminosity of LRLL 54361 was found to increase by a factor of ten in roughly 
one week every 25.34 days (Muzerolle et al. 2013).
These authors attribute this remarkable variability to pulsed accretion produced by an unseen binary companion.
The radio source appears unresolved at both wavelengths observed and its flat spectral index
is suggestive of optically-thin free-free emission. 
A peculiarity of the radio emission is that the 3.3 and 2.1 cm sources appear to be displaced by
$\sim 1''$ (see Figure 2). We consider this displacement to be significative since
our positional uncertainty is estimated to be $\sim0\rlap.{''}2$ and because all other 
unresolved sources coincide closely
in position when detected at both wavelenghts.
It is interesting to note that the 3.3 cm source is shifted towards the south-east in the same direction of
the outflow axis suggested by Muzerolle et al. (2013, see their Fig. S3), and that the position of the 2.1 cm 
source matches very well, within the positional uncertainties, the nominal position of
the centroid of the Spitzer/IRAC source 
of the region, as shown in Fig. 2.  
This suggests that the 3.3 cm emission could be tracing emission closely associated with the outflow/jet, 
while the 2.1 cm emission is probably tracing emission closer to the source driving the outflow.
This source should be monitored in the radio to test if the variability present in the
infrared extends to longer wavelengths and to understand the displacement in position at different
wavelengths.

\subsection{JVLA2}

This source coincides in position with the exciting source of the HH 211 molecular outflow
(McCaughrean et al. 1994; Gueth \& Guilloteau 1999; Palau et al. 2006). 
It was previously detected with the VLA in 1994 (Avila et al. 2001) with a flux density of
90$\pm$12 $\mu$Jy and in 2008 (Forbrich et al. 2011) with
a flux density of 87$\pm$22 $\mu$Jy, both observations at 3.5 cm. 
The flux density of 186$\pm$25 $\mu$Jy found by us for the 2013 observations
at 3.3 cm suggests that the source has increased its flux density by a factor of 2
in the period from 2008 to 2013.
The free-free jets that excite outflows and HH systems are usually steady in time.
However, evidence of variability has been presented for some of
them (e.g. Galv\'an-Madrid et al. 2004; Rodr\'\i guez et al. 2008). 
The variation detected by us is consistent with the episodic nature of
the HH~211 jet, that seems to have a period of a few decades (Lee et al. 2007).

\subsection{JVLA3}

This source is located at the center of the HH 797 outflow (McCaughrean et al. 1994; Eisl\"offel et al.
2013) and is associated with the source IC 348-SMM2 (Walawender et al. 2006). 
In the radio it is a triple source (see Figure 2), with the parameters
of its components a, b, and c given in Tables 1 and 2. The components
a and b are well separated in the 2.1 cm observations. A single radio source,
most probably the combination of components a and b, was detected
by Forbrich et al. (2011). The source MMS1 (Chen et al. 2013) also overlaps
the radio components a and b. The multiplicity of this source is
relevant because the HH 797 outflow is known to 
present apparent rotating kinematics (Pech et al. 2012). 
A double outflow system can mimic the kinematics of a rotating outflow and
our data opens the possibility that in this source the brightest radio components a and b (that are
located close to the centroid of the HH 797 outflow) may be tracing independent
jets that power nearly parallel, highly collimated outflows. 
Indeed, the 4.5 micron Spitzer image shown in Figure 1 suggests twin outflows forming the object HH 797
(see also Fig. 6 in Walawender et al. 2006), 
although the two narrow filaments seen along the source could also be tracing the limb of a single
outflow. The positive spectral indices of components a and b (Table 2)
are also consistent with a thermal jet interpretation.

The radio component 3c coincides with the millimeter source MMS2 (Chen et al. 2013), a very 
low-mass object powering a compact outflow recently detected 
by Palau et al. (2014, in preparation). 
The spectral index of this source is consistent within the noise with 
optically thick thermal free-free emission or 
with optically thin thermal dust emission (see Palau et al. 2014 for further details).

\subsection{JVLA4, 7 and 8}

These three sources lack of a counterpart at other wavelengths and their negative spectral
indices suggest optically-thin synchrotron sources, most probably faint background radio galaxies.
Following Anglada et al. (1998) we estimate that in the region imaged in Figure 1 we expect
$\sim$2 background sources above 30 $\mu$Jy at 3.3 cm. This is consistent with the number
of sources detected.
 
\subsection{JVLA5 and 6}

These radio sources coincide positionally with the young stellar objects IC 348 LRL 49
and 13 (Luhman et al. 1998), respectively. 
They are also the only sources reported here that coincide with X-ray sources, CXOUJ034357.62+320137.4  
(JVLA5) and CXOUJ034359.67+320154.1 (JVLA6), as reported in the study of
Stelzer et al. (2012). JVLA5 was detected previously at 3.5 cm with the VLA in 1994 (Avila et al. 2001), with a
flux density of 480$\pm$21 $\mu$Jy and in 2008 (Forbrich et al. 2011) with a flux density of 552$\pm$17
$\mu$Jy. These flux densities are about 3 times larger than
that reported here at 3.3 cm, implying a time-variable nature. Given these characteristics, we
propose that these two radio sources are associated with young stars with active magnetospheres,
such as those found in other regions of star formation (e.g. Dzib et al. 2013; Liu et al. 2014).
This class of extremely compact radio sources have proved to be very useful for the accurate
determination of the parallax (and distance) to several regions of star formation (e.g. Loinard et al. 2011).

\section{Conclusions}

The high sensitivity of the Jansky VLA allows the detection of a diversity of sources in
regions of star formation. The main results of our study of IC 348 SW 
can be summarized as follows.

1. We detected a total of 10 compact radio sources, determining their positions, 
flux densities and spectral indices.
Seven of these sources are new detections. 

2. One of the sources (JVLA1) is associated with the remarkable periodic time-variable infrared source
LRLL 54361 (Muzerolle et al. 2013). This detection indicates that this source can be monitored also
at radio wavelengths, helping to determine its nature.

3. Four of the sources (JVLA2, 3a, 3b, and 3c) power outflows. JVLA2 powers HH 211, while
the sources JVLA 3a and 3b appear at the center of the outflow HH 797 and
may explain its apparent rotation as due to the superposition of two nearly parallel outflows.
JVLA3c powers a compact outflow recently detected by Palau et al. (2014).

4. Two of the sources (JVLA5 and 6) are associated with infrared stars and most probably are
gyrosynchrotron sources, useful for future astrometric work.

5. Finally, three of the sources (JVLA4, 7 and 8) are most likely background extragalactic sources.

\acknowledgments

We thank James Muzerolle as well as an anonymous referee for valuable comments.
This research has made use of the SIMBAD database,
operated at CDS, Strasbourg, France.
LFR, LAZ and AP are grateful to CONACyT, Mexico and DGAPA, UNAM for their financial
support.

\clearpage


\begin{thebibliography}{}

\bibitem[Anglada et al.(1998)]{1998AJ....116.2953A} Anglada, G., 
Villuendas, E., Estalella, R., et al.\ 1998, \aj, 116, 2953 

\bibitem[Avila et al.(2001)]{2001RMxAA..37..201A} Avila, R., 
Rodr{\'{\i}}guez, L.~F., \& Curiel, S.\ 2001, \rmxaa, 37, 201 

\bibitem[Chen et al.(2013)]{2013ApJ...768..110C} Chen, X., Arce, H.~G., 
Zhang, Q., et al.\ 2013, \apj, 768, 110 

\bibitem[Dzib et al.(2013)]{2013ApJ...775...63D} Dzib, S.~A., Loinard, L., 
Mioduszewski, A.~J., et al.\ 2013, \apj, 775, 63 

\bibitem[Eisl{\"o}ffel et al.(2003)]{2003ApJ...595..259E} Eisl{\"o}ffel, 
J., Froebrich, D., Stanke, T., \& McCaughrean, M.~J.\ 2003, \apj, 595, 259 

\bibitem[Forbrich et al.(2011)]{2011ApJ...736...25F} Forbrich, J., Osten, 
R.~A., \& Wolk, S.~J.\ 2011, \apj, 736, 25 

\bibitem[Galv{\'a}n-Madrid et al.(2004)]{2004RMxAA..40...31G} 
Galv{\'a}n-Madrid, R., Avila, R., 
\& Rodr{\'{\i}}guez, L.~F.\ 2004, \rmxaa, 40, 31 

\bibitem[Gueth 
\& Guilloteau(1999)]{1999A&A...343..571G} Gueth, F., \& Guilloteau, S.\ 1999, \aap, 343, 571 

\bibitem[Herbig(1998)]{1998ApJ...497..736H} Herbig, G.~H.\ 1998, \apj, 497, 
736 

\bibitem[Hirota et al.(2008)]{2008PASJ...60...37H} Hirota, T., Bushimata, 
T., Choi, Y.~K., et al.\ 2008, \pasj, 60, 37 

\bibitem[Hirota et al.(2011)]{2011PASJ...63....1H} Hirota, T., Honma, M., 
Imai, H., et al.\ 2011, \pasj, 63, 1 

\bibitem[Lada 
\& Lada(1995)]{1995AJ....109.1682L} Lada, E.~A., \& Lada, C.~J.\ 1995, \aj, 109, 1682 

\bibitem[Lee et al.(2007)]{2007ApJ...670.1188L} Lee, C.-F., Ho, P.~T.~P., 
Palau, A., et al.\ 2007, \apj, 670, 1188 

\bibitem[Liu et al.(2014)]{2014ApJ...780..155L} Liu, H.~B., 
Galv{\'a}n-Madrid, R., Forbrich, J., et al.\ 2014, \apj, 780, 155 

\bibitem[Loinard et al.(2011)]{2011RMxAC..40..205L} Loinard, L., 
Mioduszewski, A.~J., Torres, R.~M., et al.\ 2011, Revista Mexicana de 
Astronomia y Astrofisica Conference Series, 40, 205 

\bibitem[Luhman et al.(1998)]{1998ApJ...508..347L} Luhman, K.~L., Rieke, 
G.~H., Lada, C.~J., \& Lada, E.~A.\ 1998, \apj, 508, 347 

\bibitem[McCaughrean et al.(1994)]{1994ApJ...436L.189M} McCaughrean, M.~J., 
Rayner, J.~T., \& Zinnecker, H.\ 1994, \apjl, 436, L189 

\bibitem[Muzerolle et al.(2013)]{2013Natur.493..378M} Muzerolle, J., 
Furlan, E., Flaherty, K., Balog, Z., \& Gutermuth, R.\ 2013, \nat, 493, 378 

\bibitem[Palau et al.(2006)]{2006ApJ...636L.137P} Palau, A., Ho, P.~T.~P., 
Zhang, Q., et al.\ 2006, \apjl, 636, L137 

\bibitem[Palau et al.(2014)] {} Palau, A. et al. 2014, in preparation

\bibitem[Pech et al.(2012)]{2012ApJ...751...78P} Pech, G., Zapata, L.~A., 
Loinard, L., \& Rodr{\'{\i}}guez, L.~F.\ 2012, \apj, 751, 78 

\bibitem[Preibisch(2003)]{2003A&A...401..543P} Preibisch, T.\ 2003, \aap, 401, 543 

\bibitem[Rodr{\'{\i}}guez 
\& Reipurth(1998)]{1998RMxAA..34...13R} Rodr{\'{\i}}guez, L.~F., \& Reipurth, B.\ 1998, \rmxaa, 34, 13 

\bibitem[Rodr{\'{\i}}guez et al.(2008)]{2008AJ....135.2370R} 
Rodr{\'{\i}}guez, L.~F., Moran, J.~M., Franco-Hern{\'a}ndez, R., et al.\ 
2008, \aj, 135, 2370 

\bibitem[Soker(2005)]{2005A&A...435..125S} Soker, N.\ 2005, \aap, 435, 125 

\bibitem[Stelzer et 
al.(2012)]{2012A&A...537A.135S} Stelzer, B., Preibisch, T., Alexander, F., et al.\ 2012, \aap, 537, A135 

\bibitem[Walawender et al.(2006)]{2006AJ....132..467W} Walawender, J., 
Bally, J., Kirk, H., et al.\ 2006, \aj, 132, 467 


\end{thebibliography}
\end{document}